\documentclass[
  journal=large,
  manuscript=,
  year=2020,
  volume=37,
]{cup-journal}
\usepackage{amsmath,amssymb,amsfonts}
\usepackage{bm}
\usepackage{url}
\usepackage{subfig}
\usepackage{graphicx}
\usepackage{comment}
\usepackage{floatrow}
\usepackage{caption}
\usepackage{siunitx}
\usepackage{microtype}
\usepackage{booktabs}
\def\IN{\rm IN}
\def\OUT{\rm OUT}
\def\vec#1{\bm #1}
\DeclareMathOperator{\Tr}{\mathrm{Tr}}

\newcommand{\ket}[1]{\ensuremath{\left| #1 \right\rangle}}
\newcommand{\bra}[1]{\ensuremath{\left\langle #1 \right|}}

\usepackage[hidelinks]{hyperref}
\usepackage{color}

\title{Robustness of Energy Landscape Control to Dephasing}

\author{Sean P.\ O'Neil}
\affiliation{Department of Electrical and Computer Engineering, University of Southern California, Los Angeles, CA 90089} 
\email[S O'Neil]{seanonei@usc.edu}

\author{Frank C.\ Langbein}
\affiliation{School of Computer Science and Informatics, Cardiff University, Cardiff CF24 4AG, UK} 
\email[F Langbein]{frank@langbein.org}

\author{Edmond Jonckheere}
\affiliation{Department of Electrical and Computer Engineering, University of Southern California, Los Angeles, CA 90089}
\email[E Jonckheere]{jonckhee@usc.edu}

\author{S.\ Shermer}
\affiliation{Faculty of Science and Engineering, Physics, Singleton Park, Swansea, SA2 8PP, UK}
\email[S Shermer]{s.m.shermer@gmail.com}


\begin{document}

\begin{abstract}
As shown in previous work, in some cases closed quantum systems exhibit a non-conventional absence of trade-off between performance and robustness in the sense that controllers with the highest fidelity can also provide the best robustness to parameter uncertainty. As the dephasing induced by the interaction of the system with the environment guides the evolution to a more classically mixed state, it is worth investigating what effect the introduction of dephasing has on the relationship between performance and robustness. In this paper we analyze the robustness of the fidelity error, as measured by the logarithmic sensitivity function, to dephasing processes. We show that introduction of dephasing as a perturbation to the nominal unitary dynamics requires a modification of the log-sensitivity formulation used to measure robustness about an uncertain parameter with non-zero nominal value used in previous work. We consider controllers optimized for a number of target objectives ranging from fidelity under coherent evolution to fidelity under dephasing dynamics to determine the extent to which optimizing for a specific regime has desirable effects in terms of robustness. Our analysis is based on two independent computations of the log-sensitivity: a statistical Monte Carlo approach and an analytic calculation. We show that despite the different log-sensitivity calculations employed in this study, both demonstrate that the log-sensitivity of the fidelity error to dephasing results in a conventional trade-off between performance and robustness.
\end{abstract}

\section{Introduction} \label{intro}

The advent of quantum technology and promise of applications ranging from quantum computing to quantum sensing has resulted in strong interest in a range of quantum systems. In particular coupled spin systems, or spin networks for short, show potential as simple prototypes on the path to scaling to more complex systems~\parencite{Awschalom2013}. As control plays a fundamental role in the translation of physical phenomena into technology, the development and implementation of effective control schemes for quantum systems are essential to harness their technological potential~\parencite{Glaser2015}. Coupled with the design of controllers for quantum systems, tools to assess and guarantee robustness of these controllers to the effect of the environment are essential to realizing the benefits of quantum technology for high fidelity medical imaging, operation of quantum gates, and quantum computing~\parencite{bio-imaging,Koch_2022,quantum_dots}.  

A paradigm for quantum control based on energy landscape shaping has been proposed and applied to derive feedback control laws for selective transfer of excitations between different nodes in a spin network~\parencite{CDC2018,time_optimal,Edmond_IEEE_AC}. The controllers $D(\ket{\IN}, \ket{\OUT})$ for this scheme are designed to maximize the fidelity $\left|\bra{\OUT}U(T)\ket{\IN}\right|^2$ of transfer from an input state $\ket{\IN}$ to an output state $\ket{\OUT}$ at a specified readout time $T$. $U(T)$ is the unitary time-evolution operator of the system, which depends on static fields to shift the energy levels of the system~\parencite{time_optimal, DataSet1}. These optimal controllers $D$ are selective in that no input other than $\ket{\IN}$ can drive the system to $\ket{\OUT}$ at time $T$.

Although this quantum control problem can be formulated as a linear time-invariant (LTI) control system with state feedback, there are numerous differences between this quantum control problem and a classical tracking problem. The unitary evolution of a closed quantum system is characterized by persistent oscillations. As such, the system is not stable in a classical sense, and the target states are not attractive or asymptotically stable steady states.  
Lack of stability might be expected to bode poorly for robustness \CUPTWOCOL
but this is not necessarily always the case. In~\parencite{Edmond_IEEE_AC} it was shown that controllers achieving perfect state transfer also have vanishing sensitivity with respect to perturbations to the coupling strengths of the drift Hamiltonian. At the same time, statistical analysis of a set of optimal energy landscape controllers for uniformly coupled spin rings, ranging in size from $3$ to $20$ spins, found that under certain conditions, a concordant relationship between the error and the log-sensitivity is possible for controllers that achieve low but non-vanishing errors~\parencite{statistical_control}. 
 
Our goal in this paper is to investigate whether a non-conventional lack of trade-off between robustness and performance is observed in simple spin-rings evolving under dephasing dynamics or whether a conventional trade-off exists. We assess robustness through the log-sensitivity, calculated both analytically and via kernel density estimation (KDE). We also consider controllers optimized to provide maximum fidelity under different assumptions on the system-environment interaction. In particular, we consider controllers optimized for maximum fidelity under unitary evolution, those optimized for fidelity with dephasing introduced, and those optimized for a linear combination of fidelity under unitary dynamics and steady-state fidelity. To understand this effect of decoherence on the controller design we will focus on the intermediate regime where coherent dynamics play a significant role but are modified by dephasing as a result of weak interactions with the environment. The strongly dissipative regime where asymptotic stability can be recovered~\parencite{Schirmer2010} and exploited to design backaction-based stabilization schemes~\parencite{Ticozzi2010, Motzoi2016} has been considered in other work~\parencite{Schirmer2022}.

While the non-conventional absence of a trade-off between robustness and performance observed in some cases under coherent dynamics may carry over to systems subject to decoherence, the addition of decoherence alters the dynamics significantly. Pure dephasing, in particular, results in quantum superposition states converging to classical mixed states. It is thus reasonable to expect a more classical, in a control-theoretic sense, behavior for systems subject to decoherence.

In Section~\ref{s:spin-systems} we introduce the theory of coupled spin systems and their evolution under decoherence in the single excitation subspace. In Section~\ref{design} we introduce the control objectives in terms of maximization of the transfer fidelity and the objective functions for optimal controller design in the different regimes. Next, in Section~\ref{lti}, we introduce an LTI form of the dynamical equations amenable to robustness analysis.  In Section~\ref{robust} we provide the pair of methods (statistical and analytic) used to gauge the sensitivity and robustness of the controllers. In Section~\ref{analysis} we present results detailing the level of concordance between the two log-sensitivity calculations, the degree to which the robustness properties of the controllers agree with the trade-offs from classical control theory, and explore the effect of specific controller types on the fidelity and observed robustness properties. Finally, we conclude with Section~\ref{conclusion}. 

\section{Methods} \label{method}

In this section we outline the equations governing excitation transfer for spin rings in the single excitation subspace, describe the control objectives and optimization scheme used to develop the controllers, and detail the methods used to assess the robustness of these controllers to perturbations in the form of dephasing.

\subsection{Coupled Spin Dynamics --- Single Excitation Subspace} \label{s:spin-systems}

We consider a ring of $N$ spin-$1/2$ particles with nearest-neighbor coupling in the subspace of the state space where the total number of excitations is one, which consists of states where one spin is in an excited state and $N-1$ spins remain in the ground state, and superpositions of such states. As detailed in~\parencite{time_optimal, Edmond_IEEE_AC, data_set_1_analysis} the Hamiltonian for this spintronic network in the single-excitation subspace is represented as
\begin{equation}\label{e:Hamiltonian}
H_D = \begin{pmatrix}
  D_1      & J_{1,2} & 0       & \ldots & 0         & J_{1,N} \\
  J_{1,2}  & D_2     & J_{2,3} &        & 0         & 0\\
  0        & J_{2,3} & D_3     &        & 0         & 0\\
  \vdots   &         & \ddots  & \ddots & \ddots    & \vdots\\
  0        &  0      & 0       &        & D_{N-1}   & J_{N-1,N}\\
  J_{1,N}  & 0       & 0       & \ldots & J_{N-1,N} & D_{N}
  \end{pmatrix}
\end{equation}
in the basis where each natural basis vector represents the excitation localized at spin $\ket{n}$. The terms $J_{k,\ell}$ represent the coupling between spins $\ket{k}$ and $\ket{\ell}$ and are all assumed equal. The terms $D_n$ are scalar values of the time-invariant control fields applied to shape the energy landscape. This single excitation subspace model is a simplification of the model for a system with any number of excited spins up to $N$. Specifically, this is the subspace that results by retaining only those eigenvectors with eigenvalue $1$ for the total spin operator $S_N=\frac{1}{2}\sum_{k = 1}^{N}(I+Z_k)$~\parencite{Joel_2013}.

Assuming weak interaction with the environment, the dynamics of the system are described by the Lindblad differential equation: 
\begin{equation} \label{lindblad}
  \dot{\rho}(t) = -\tfrac{j}{\hbar} [H_D,\rho(t)] + L_D(\rho(t)),
\end{equation}
where $\hbar$ is the reduced Planck constant, $H_D$ the Hamiltonian defined above and $L_D$ is a Lindblad super-operator
\begin{equation}\label{e:V}
  L_D(\rho) =  {V_{D}}^{\dagger} \rho V_D - \tfrac{1}{2} ({V_{D}}^\dag V_D \rho + \rho {V_D}^\dag V_D).
\end{equation}
By setting $V_D=0$ we recover the usual Hamiltonian dynamics considered in previous work~\parencite{Edmond_IEEE_AC,data_set_1_analysis}. Here, we study systems subject to decoherence that can be modeled as dephasing in the Hamiltonian basis. This is a common model for weak decoherence, described by a Lindblad operator $L_D$ of dephasing type, given by a Hermitian dephasing operator that commutes with the system Hamiltonian,
\begin{equation}
  V_D = {V_D}^\dag \text{ where }[H_D,V_D] = 0.
\end{equation}
The subscript $D$ in the Lindbladian indicates a dependence on the control as, strictly speaking, decoherence in the weak coupling limit depends on the total Hamiltonian, and hence on the control~\parencite{domenico_CDC, singular_vs_weak_coupling}.  Although this is a simple decoherence model, it is closer to the master equation in the weak coupling limit developed in~\parencite{singular_vs_weak_coupling} as it appears at first glance. For a Hermitian $V_D$, it is easily verified that the Lindblad superoperator simplifies to
\begin{equation}\label{e:QME2}
  L_D(\rho) = - \tfrac{1}{2} [V_D, [V_D, \rho]].
\end{equation}
As $H_D$ and $V_D$ commute for the dephasing model, they are simultaneously diagonalizable and there exists a set of projectors $\{\Pi_k\}_{k}$ onto the (orthogonal) simultaneous eigenspaces of $H_D$ and $V_D$ such that $\sum_{k}\Pi_k=I_{\mathbb{C}^{N}}$ is a resolution of the identity on the Hilbert space $\mathbb{C}^{N}$ of the single excitation subspace and
\begin{equation*}
  H_D = \sum_{k} \lambda_k \Pi_k, \quad
  V_D = \sum_{k} c_k \Pi_k,
\end{equation*}
where $\lambda_k$ and $c_k$ are the real eigenvalues of $H_D$ and $V_D$, respectively.

Pre- and post-multiplying the master Equation~\eqref{lindblad} with Lindblad term~\eqref{e:V} by $\Pi_k$ and $\Pi_\ell$ yields
\begin{equation} \label{e:rho_k_ell}
  \Pi_k\dot{\rho}(t) \Pi_\ell = (-j \omega_{k \ell} - \gamma_{k \ell}) \Pi_k \rho(t) \Pi_\ell,
\end{equation}
where $\omega_{k \ell}= \tfrac{1}{\hbar}(\lambda_k - \lambda_\ell)$ and $\gamma_{k\ell} = \tfrac{1}{2\hbar}(c_k-c_\ell)^2 \ge 0$. The solution for the projection $\Pi_k \rho(t) \Pi_\ell$ is 
\begin{equation*}
  \Pi_k\varrho(t)\Pi_\ell = e^{-t(j \omega_{k\ell} + \gamma_{k\ell})} \Pi_k \varrho_0 \Pi_\ell.
\end{equation*}
Since $\sum_k \Pi_k = I_{\mathbb{C}^N}$, the full solution is
\begin{equation}\label{e:varrho_of_t_solution}
 \varrho(t) = \sum_{k,\ell} e^{-t(j \omega_{k\ell} + \gamma_{k\ell})} \Pi_k \varrho_0 \Pi_\ell.
\end{equation}

\subsection{Control Objectives and Controller Design} \label{design}

In this section we define the control objective as the transfer fidelity and discuss differing conditions under which we seek to maximize this measure. These varying conditions manifest as distinct sets of controllers aimed at optimizing the fidelity under differing conditions.  

\subsubsection{Transfer Fidelity}\label{ss:fidelity}

Following the framework adopted in earlier work~\parencite{time_optimal,Edmond_IEEE_AC}, we seek static controllers that map an input state to a desired output state by shaping the energy landscape of the system. Specifically, our design objective is to find a controller that steers the dynamics to maximize the transfer fidelity of an excitation at an initial node of the network, $\ket{\IN}$, to an output node $\ket{\OUT}$ at a specific read-out time $T$. If the output state is a pure state, this target state can be represented as $\rho_{\OUT} = \ket{\OUT}\bra{\OUT}$. We then evaluate the fidelity of the state $\rho(t)$ at time $T$ in terms of the overlap with $\rho_{\OUT}$ as
\begin{equation} \label{e:transfer_fid}
  F[\rho(T)] =  \Tr [\rho_{\OUT}\rho(T)].
\end{equation}
The maximum transfer fidelity, $F_{\max}=1$, is attained when $\rho(T)=\rho_{\OUT}$ as
\begin{equation}
  F[\rho(t)] \leq F[\rho_{\OUT}] = \Tr[\rho_{\OUT}^2] = 1.
\end{equation}
The fidelity error $e(T)$ at the readout time $T$ is therefore given by $e(T) = 1-F[\rho(T)]$. We thus seek controllers that maximize this transfer fidelity (equivalently minimize the fidelity error) defined in~\eqref{e:transfer_fid}, where $\rho(T)$ is the solution of Eq.~\eqref{lindblad} with $\rho(0) = \rho_0 =\ket{\IN}\bra{\IN}$.

\subsubsection{Optimal Controller Design}\label{ss:controller_design}

For the energy landscape control paradigm, finding a controller is equivalent to finding an ordered $N$-tuple of control parameters $\begin{bmatrix} D_1, \ D_2, \hdots, \ D_N \end{bmatrix}^T$ and a time $T$ that maximizes the transfer fidelity for a system evolving under the total Hamiltonian $H_D$, according to Eq.~\eqref{lindblad}. If the decoherence process is known precisely then it is straightforward to optimize the control objective by evolving the system according to Eq.~\eqref{lindblad} and evaluating the fidelity. However, the exact dephasing rates for a given system are often not precisely known~\parencite{deph_rates}. We thus consider three different scenarios for optimal controller synthesis:   
\begin{enumerate}
\item Optimize the transfer fidelity under unitary dynamics.
\item Optimize a convex combination of unitary transfer and asymptotic transfer fidelity.
\item Optimize the fidelity averaged over a sampling of decoherence processes.
\end{enumerate}

Option~1 is a reasonable choice if decoherence is a weak perturbation to the Hamiltonian dynamics. Optimizing solely for asymptotic transfer fidelity may be a reasonable choice if the decoherence is so strong that the system is likely to reach a steady state before the transfer is complete. However, in the intermediate regime, when we are unsure of the dephasing rates, optimizing for Options~2 and~3 is more practical.

\subsubsection{Optimization under Unitary Dynamics}

Optimization of Option~1 has been considered in previous work ~\parencite{time_optimal,Edmond_IEEE_AC,DataSet1}. The optimization problem in the other cases can be solved similarly, using standard optimization algorithms with suitable modification of the objective functional. Despite the complex optimization landscape, we have found that the L-BFGS (Limited memory Broyden–Fletcher–Goldfarb–Shanno) quasi-Newton algorithm with restarts using randomly selected initial values in a sufficiently large domain based on stratified sampling works well for all options~\parencite{time_optimal,DataSet2_code}.

\subsubsection{Optimization of Coherent and Asymptotic Transfer Fidelity} \label{ss:simult_optim}

To simultaneously optimize coherent and asymptotic transfer, we define an objective function that is a weighted average of both, e.g.,
\begin{equation}
  \alpha \Tr[\rho(T) \rho_{\OUT}] + \left(1-\alpha\right) \Tr[\rho_{\infty}\rho_{\OUT}],
\end{equation}
where $\rho(t)=U(t)\rho_0 U(t)^\dag$ is the initial state propagated by unitary evolution and $\rho_\infty$ is the steady state of decoherent evolution. Both $\rho(T)$ and $\rho_{\infty}$ are efficiently calculated from~\eqref{e:varrho_of_t_solution} by setting $t=T$, $\gamma_{k\ell}=0$ for coherent evolution and $t=\infty$, $k=\ell$ for the decoherent steady state, respectively.  

To maximize this weighted average fidelity, a controller must be superoptimal. Specifically, it must enable perfect state transfer from $\rho_{0}$ to $\rho_{\OUT}$ at time $t=T$. Simultaneously, it must maximize the overlap of the steady state $\rho_{\infty}$ with the target state $\rho_{\OUT}$. To see that such controllers exist in principle, consider a controller that achieves maximum asymptotic transfer by rendering the input and output states orthogonal superposition states of the form $\ket{\IN} = \frac{1}{\sqrt{2}} \left( \ket{e_1} + \ket{e_2} \right)$ and $\ket{\OUT} = \frac{1}{\sqrt{2}} \left( \ket{e_1} - \ket{e_2} \right)$ in the eigenbasis of the Hamiltonian, $H_D = \sum_{k=1}^N E_k \ket{e_k}\bra{e_k}$. We refer to these states as orthogonal pairs. Here, $E_k$ and $\ket{e_k}$ represent the energy eigenvectors and eigenvalues of $H_D$. Since $\rho_0$ and $\rho_{\OUT}$ only involve the first two eigenvectors of $H_D$, we can restrict ourselves to considering the representation on this subspace,
\begin{equation}
\rho_{0} = \frac{1}{\sqrt{2}} \begin{bmatrix} 1 & 1 \\ 1 & 1 \end{bmatrix}, \quad 
\rho_{\OUT} = \frac{1}{\sqrt{2}} \begin{bmatrix} 1 & -1 \\ -1 & 1 \end{bmatrix}. 
\end{equation}
The general state evolves as 
\begin{equation}
\rho(t) = \frac{1}{\sqrt{2}} \begin{bmatrix} 1 & e^{-t(j \omega_{12} + \gamma_{12})} \\ e^{-t(j \omega_{21} + \gamma_{21})} & 1 \end{bmatrix}.
\end{equation}
In terms of the first objective (fidelity under unitary dynamics), $\gamma_{12} = \gamma_{21} = 0$, and at a time $T = \frac{ (2n + 1) \pi}{\omega_{12}}$ for $n \in \mathbb{Z}$ the controller achieves perfect state transfer or $\rho(t) = \rho_{\OUT}$, maximizing the first half of the objective. In terms of the asymptotic component of the objective, with $\gamma_{12} = \gamma_{21} > 0$,
\begin{equation}
    \lim_{t \to \infty} \rho(t) = \frac{1}{\sqrt{2}} \begin{bmatrix} 1 & 0 \\ 0 & 1 \end{bmatrix}.
\end{equation}
And so $\Tr\left[ \rho_{\infty}\rho_{\OUT} \right] = \frac{1}{2}$, maximizing the possible overlap. 

\subsubsection{Decoherence-averaged Optimization}\label{ss:dec_optim}

Optimization of the transfer fidelity, averaged over many dephasing processes, is in principle also straightforward. For a given initial state $\rho_0$ and controller $D$, the output state $\rho^{(D,S)}(T)$ subject to a dephasing process $S$ is calculated according to~\eqref{e:varrho_of_t_solution} and transfer fidelity from~\eqref{e:transfer_fid}. From this the average transfer fidelity can be computed by taking the mean of the transfer fidelity over all decoherence processes.

The computational overhead of the average fidelity evaluation, and thus the optimization as a whole, depends linearly on the number of decoherence processes averaged over. Efficient sampling of the possible decoherence processes to minimize the number of required decoherence processes and avoid sampling bias is therefore important.

Decoherence in the form of dephasing in the Hamiltonian basis is modeled by sampling the space of pure dephasing processes. We generate a large set of $N \times N$ lower triangular matrices with entries $\Gamma_{k\ell}^{(S)}$ in $[0,1]$. Here, $N$ is the system dimension given by the number of qubits in the network. To ensure even sampling of the whole space, the entries of the triangular matrices are drawn from a Sobol sequence for low-discrepancy sampling, thus allowing an even covering of the sample space~\parencite{Burhenne_2011}. A set of at least $10,000$ dephasing operators is then generated by eliminating all trial dephasing matrices that violate the complete-positivity physical constraints for evolution of an open quantum system~\parencite{Gorini_1977,PhysRevA.86.012121}. We further normalize each dephasing matrix $\Gamma_{k\ell}^{(S)}$ as
\begin{equation}\label{eq: gamma}
  \bar{\Gamma}_{k\ell}^{(S)} = \Gamma_{k\ell}^{(S)} \left/ \sum_{1 < k \le N, 1 \le \ell < k} |\Gamma_{k\ell}^{(S)}|\right..
\end{equation}
We then use $1,000$ of these dephasing operators in each optimization run for fidelity with dephasing dynamics. 

\subsection{LTI Formulation of the State Equation} \label{lti}

To facilitate the following analysis we reformulate the Lindblad equation~\eqref{lindblad} and its solution~\eqref{e:varrho_of_t_solution} through expansion by a set of $N^2$ basis matrices for Hermitian operators on the space $\mathbb{C}^N$~\parencite{Altafini2012}. The result of this vectorization process is to transform the master equation~\eqref{lindblad} to the linear time-invariant form
\begin{equation}\label{eq: lti_sol}
    \dot{r}(t) = (A + L)r(t) 
\end{equation}
where $r(t) \in \mathbb{R}^{N^2}$ is the vectorized version of the density matrix in the Hamiltonian basis, $A \in \mathbb{R}^{N^2 \times N^2}$ is the matrix representation of the Liouville superoperator which determines the unitary evolution of the system, and $L \in \mathbb{R}^{N^2 \times N^2}$ is the matrix representation of the Lindblad superoperator. To see this, consider the decomposition of the controlled Hamiltonian $H_D = U \Lambda U^{\dagger}$ where $U \in \mathit{U}(N)$ and $\Lambda \in \mathbb{R}^{N \times N}$ is the diagonal matrix of real eigenvalues of $H_D$. Pre-multiplying~\eqref{lindblad} by $U^{\dagger}$, post-multiplying by $U$, and noting that $UU^{\dagger} = U^{\dagger}U = I$, we have
\begin{equation}\label{eq: diag}
\hbar\dot{\tilde{\rho}}(t) = j\Lambda \tilde{\rho}(t) + j \tilde{\rho}(t) \Lambda + C\tilde{\rho}(t)C -\tfrac{1}{2}C^2\tilde{\rho}(t) - \tfrac{1}{2} \tilde{\rho}(t) C^2
\end{equation}
where $\tilde{\rho}(t) = U^{\dagger}\rho(t)U$ is the representation of the density matrix in the Hamiltonian basis, and $C = U^{\dagger} V_D U$ is a diagonal matrix of the $N$ scalar $c_k$'s from the decomposition $V_D = \sum_{k}c_k \Pi_k$. 

We choose the $N^2-1$ generalized Pauli matrices~\parencite{Bertlmann_2008} complemented by $I_N$ to form an orthonormal basis for the Hermitian operators on $\mathbb{C}^{N}$ which we designate as $\{\sigma_n\}$. The orthonormality conditions are expressed as $\Tr\left(\sigma_m,\sigma_n\right) = \delta_{mn}$. Expansion of~\eqref{eq: diag} in terms of the basis $\{\sigma_n\}$, yields the following~\parencite{neat_formula}:
\begin{subequations}
  \begin{align}
    r_k(t) &= \Tr \left( \tilde{\rho}(t) \sigma_k \right)\\
    A_{k \ell} &= \Tr\left(\frac{j}{\hbar} \Lambda  [\sigma_k,\sigma_{\ell}]\right), \label{eq: mapping-A}\\
    L_{k\ell} &= \frac{1}{\hbar}\Tr(C \sigma_k C \sigma_{\ell}) - \frac{1}{2 \hbar} \Tr(C^2 \{\sigma_k,\sigma_\ell\}) \label{eq: mapping-L}.
  \end{align}
\end{subequations}
Here $\left[ \cdot, \cdot \right]$ is the matrix commutator and $\left\{ \cdot,\cdot \right\}$ is the anti-commutator. The solution to~\eqref{eq: lti_sol} is 
\begin{equation} \label{eq: linear_solution}
r(t) = e^{t(A + L)}r_0
\end{equation}
where $r_0 \in \mathbb{R}^{N^2}$ is the vectorized version of $\rho_0$ with components given by $r_{0k} = \Tr(\rho_0 \sigma_k)$. 

Before proceeding, we make a few observations. Firstly, as $C$ and $\Lambda$ are diagonal and commute, then so do $A$ and $L$ which simplifies calculation of the log-sensitivity. Secondly, given the requirement that $\Tr(\rho(t)) = 1\;\forall t$, we see that $r_{N^2}(t) = \frac{1}{\sqrt{N}}$, which implies $\dot{r}_{N^2} = 0$. This guarantees the existence of at least one non-zero eigenvalue of the dynamical equation~\eqref{eq: lti_sol} and, similarly, the existence of a subspace along which the trajectory is constant (a steady state).

Additionally, if the eigenvalues of $A + L$ not equal to zero are distinct, and the dephasing process is characterized by $N$ distinct jump operators not equal to zero, then $A + L$ has $N$ zero eigenvalues corresponding to $k = \ell$, consistent with the constant populations on the main diagonal of $\rho(t)$ in~\eqref{e:varrho_of_t_solution}. We return to this linear time-invariant formulation to assess robustness in Section~\ref{robust}.   

\subsection{Robustness Assessment} \label{robust}

Following~\parencite{CDC2018,statistical_control,oneil_2022}, we assess the robustness of the implemented controllers to perturbations to the nominal system through the logarithmic sensitivity of the fidelity error, or log-sensitivity for short.

We consider our nominal system as the closed system evolving under unitary dynamics with a nominal trajectory given by~\eqref{e:varrho_of_t_solution} with $\gamma_{k \ell} = 0$ or the vectorized form~\eqref{eq: lti_sol} with $L = \mathbf{0}_{N^2 \times N^2}$. We then consider perturbations of this nominal trajectory due to the introduction of dephasing. We represent the dephasing process as the matrix $S_\mu \in \mathbb{C}^{N^2 \times N^2}$ where $\mu$ is used to index distinct dephasing processes. Even though $S_\mu$ is an operator on the Hilbert space over $\mathbb{C}^N$, it consists of strictly real elements $\bar{\Gamma}^{(S_{\mu})}_{k\ell} = \gamma^{(S_{\mu})}_{k \ell}$ as defined in~\eqref{eq: gamma}. We will consider a set of $1000$ dephasing operators in each trial, indexed as $\mu \in \{1,2,\hdots,1000\}$. Furthermore, we introduce a parameter $\delta \in \left[0,1\right]$ to modulate the strength of the dephasing process for each $S_{\mu}$. We then denote the perturbed trajectory in the density matrix formalism as
\begin{equation}\label{eq: perturbed_rho}
\tilde{\rho}(t;S_{\mu},\delta) = \sum_{k,\ell}e^{-t(j\omega_{kl} - \delta \gamma^{(S_{\mu})}_{k\ell})} \Pi_{k}\rho_0 \Pi_{\ell}
\end{equation}
or in the LTI formulation as 
\begin{equation}\label{eq: perturbed_r}
    \tilde{r}(t;S_{\mu},\delta) = e^{t(A + \delta S_{\mu})}r_{0}. 
\end{equation} 
The fidelity error for the perturbed density matrix of~\eqref{eq: perturbed_rho} is then given as 
\begin{equation}\label{eq: perturbed_err_rho}
    \tilde{e}(T;S_{\mu},\delta) = 1 - \Tr[\tilde{\rho}(T;S_{\mu},\delta)\rho_{\OUT}]
\end{equation}
or 
\begin{equation}\label{eq: perturbed_err_r}
    \tilde{e}(T;S_{\mu},\delta) = 1 - \vec{c}e^{(A + \delta S_{\mu})T}r_0
\end{equation}
where $\vec{c}$ is the $N^2 \times 1$ row vector corresponding to the transpose vectorized representation of $\rho_{OUT}$.
 
We are now in a position to assess the robustness of $\tilde{e}(T;S_{\mu},\delta)$ by measuring the effect of the dephasing perturbation through the log-sensitivity~\parencite{oneil_2022}. We rely on the log-sensitivity based on its widespread use and relation to fundamental limitations of classical control~\parencite{statistical_control,dorf}. Specific to the excitation transfer problem, we quantify performance as high fidelity $F[\rho(T)]$ (equivalently low error $1 - F[\rho(T)] = e(T)$) and measure robustness as the size of the differential change in the performance induced by an external perturbation or $\frac{\partial e(T)}{\partial \delta}$. In terms of fundamental limitations, we expect that those controllers which exhibit the best performance will display the greatest differential log-sensitivity (the least robustness) and vice-versa for those controllers exhibiting low performance.

Next, in order to better gauge a normalized percentage change in $\tilde{e}(T)$ for a given $S_\mu$, we consider a logarithmic sensitivity of the form $\frac{\partial \ln(\tilde{e}(T)}{\partial \delta}$. We calculate the log-sensitivity of the fidelity error to the perturbation $S_{\mu}$ as
\begin{equation}\label{eq: log-sens}
    s(S_{\mu},T) =\left. \frac{\partial \ln(\tilde{e}(T;S_\mu,\delta))}{\partial \delta} \right|_{\delta = 0} = \left. \frac{1}{e(T)}\frac{\partial \tilde{e}(T;S_{\mu},\delta)}{\partial \delta} \right|_{\delta = 0}.
\end{equation}
Note that we depart from the definition of log-sensitivity used in~\parencite{oneil_2022}, as we see two distinct cases when applying the log-sensitivity. In the first case there are no changes in the inertia of the system matrix when parameters drift about their nominal values. By inertia of a system matrix, we mean the number of eigenvalues $\lambda_k$ with $\Re{\lambda_k} = 0$, $\Re{\lambda_k} < 0$, and $\Re{\lambda_k}>0$, characterizing the response of the system. This first case is the basis for the analysis in~\parencite{oneil_2022} and includes the uncertain couplings $J$ as in~\parencite{statistical_control}. In such cases, the genuine log-sensitivity $\partial \ln(e(T)) / \partial \ln(\xi) \rvert_{\xi = \xi_0}$ provides a meaningful, dimensionless measure of sensitivity to the uncertain parameter $\xi$ with nominal value $\xi_0$. In the more complicated second case there are changes of inertia around the nominal parameter values. The present case falls in this category as the nominal decoherence rate is $\delta_{0}=0$. Introduction of the perturbation induces a bifurcation in the system dynamics from unitary evolution to decoherent evolution. In this case, evaluating the log-sensitivity as $\partial \ln(e(T)) / \partial \ln(\delta) \rvert_{\delta=\delta_0}$ yields a zero value for all controllers. This requires a revision of the log-sensitivity as $(\partial e(T) / \partial \delta) \cdot (1/e(T)) \rvert_{\delta=0}$ to obtain a meaningful log-sensitivity. 
Regardless, $s(S_{\mu},T)$ as defined in~\eqref{eq: log-sens} provides a percentage change in the error with respect to the introduction of dephasing. Put differently, the value of $s(S_{\mu},T)\delta$ for non-vanishing $\delta$ provides a means to compare the effect of decoherence process $S_\mu$ of strength $\delta$ across different controllers.

\subsection{Kernel Density Estimate Approach}\label{ss: mc}

As shown in~\parencite{CDC2018}, evaluating the log-sensitivity of the error numerically over a large number of perturbation provides one method of assessing robustness. In the current study, we use this as the first approach to calculating the log-sensitivity. Specifically, we consider spin rings of size $N=5$ and $N=6$. For each ring, we consider transfers from spin $\ket{\IN} = \ket{1}$ to $\ket{\OUT} = \ket{2,3}$ for $N=5$ and $\ket{\OUT} = \ket{2,3,4}$ for $N = 6$. For each transfer we select the best, as measured by highest nominal fidelity, $100$ controllers from the three optimization categories described in Section~\ref{ss:controller_design}: fidelity under unitary dynamics, fidelity under dephasing, and unitary transfer combined with asymptotic fidelity. For brevity, we refer to these optimization options as fidelity, dephasing, and overlap, respectively. For each controller, we select $1000$ dephasing operators generated by the process described in Section~\ref{ss:dec_optim}. For each dephasing process, we consider a perturbation $\delta \in [0,1]$ quantized into $1001$ points with a uniform interval of $0.001$. 

With this set-up, we calculate $\tilde{e}(T;S_{\mu},\delta)$ for each controller across the entire population of $S_{\mu}$ and $\delta$ by~\eqref{eq: perturbed_rho} and~\eqref{eq: perturbed_err_rho}. The result is an $1000 \times 1001$ element array of error results arranged by dephasing process along the rows and perturbation strength along the columns. This array serves as the kernel for the MATLAB function \texttt{ksdensity} to compute a kernel density estimate of the error to dephasing for a given strength~\parencite{Silverman_1986}. The bandwidth used in the calculation is $h = 3.5 \sigma n^{-1/3}$ as derived in~\parencite{scott_1979} where $\sigma$ is the standard deviation of the samples in each error array noted above and $n$ is the number of samples. For each step in $\delta$ we also calculate the mean and variance of the error across the $1000$ dephasing processes. These $1001$ samples of the mean error over all dephasing operators serves as input to the MATLAB function \texttt{fit} with option \texttt{'smoothingspline'} to produce a functional representation of the mean error designated as $\hat{e}(T;S,\delta)$ where we drop the subscript $\mu$ on the dephasing operator to indicate that the averaging process in the density estimation has already been taken  into account. We choose a smoothing spline over a cubic fit to provide the greatest degree of freedom while providing a functional fit that minimizes any distortion in the data~\parencite{splines}. A numeric differentiation of $\hat{e}(T;S,\delta)$, evaluated at the point where $\delta = 0$, then provides an estimate of the differential sensitivity. We then have
\begin{equation}
    s_{k}(S,T) = \frac{1}{e(T)}\left. \frac{\partial \hat{e}(T,S,\delta)}{\partial \delta} \right|_{\delta = 0}
\end{equation} 
where $e(T) = \hat{e}(T;S,0)$ is the nominal error. The subscript $k$ indicates the value of the log-sensitivity is calculated from the density estimate of the mean error. 

\begin{figure*}[ht]
\centering
\includegraphics[width=\textwidth]{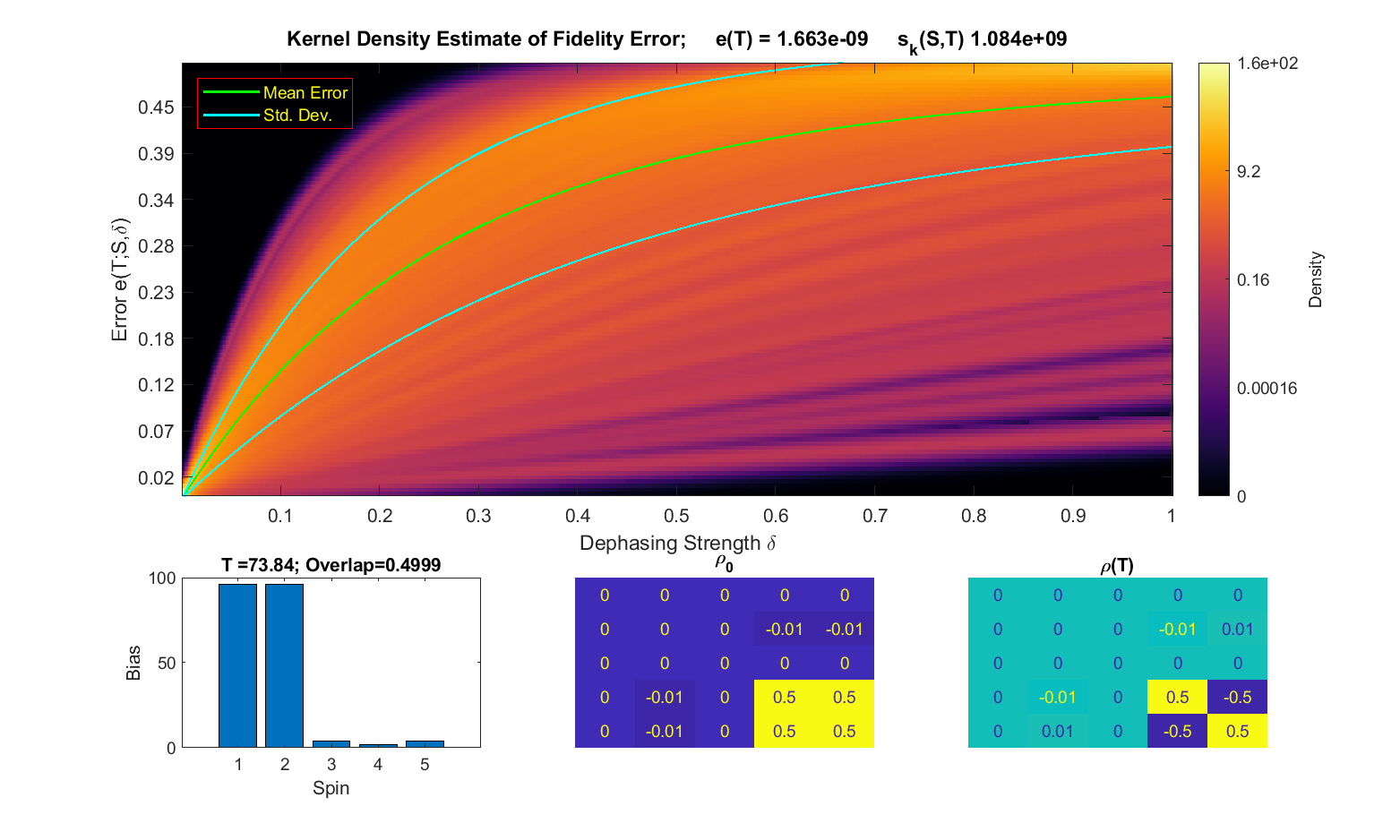}
\caption{Top: heat map of KDE-based fidelity error distribution as a function of the dephasing strength for a $0 \rightarrow 1$ transfer in a $5$-ring with dephasing-optimized controller.  The green lines indicate the mean and standard deviation of the distribution. The slope of the mean error as a function of the decoherence strength $\delta$ is used to estimate the sensitivity in the limit $\delta\to 0$, which provides the numerical estimate $s_{k}(S,T)$ of the log-sensitivity. Bottom left: values of the biases of the energy landscape controller, with transfer time $T$ and steady-state overlap indicated in the title. Bottom center and right: heat map of the initial state $\rho_0$ and final state $\rho(T)$ with respect to an eigenbasis of the controlled Hamiltonian. Notice that in this example, the initial and final state form an orthogonal pair in this basis. }\label{fig:summary}
\end{figure*}

As an example of the output of this kernel density estimation, Figure~\ref{fig:summary} displays a heatmap visualization of the error versus decoherence strength in the upper pane. The slope of the green line at the point where $\delta = 0$ provides the estimate of the differential sensitivity used in the log-sensitivity calculation. The repository located at~\parencite{DataSet2_results} contains the entire collection of these figures for all controllers and optimization options. 

\subsection{Analytic Calculation}

The structure of the matrices $A$ and $S_{\mu}$ greatly simplifies the calculation of the log-sensitivity. Based on the fidelity error defined in~\eqref{eq: perturbed_r} and noting that $A$ and $S_{\mu}$ commute we have that $\tilde{e}(T;S_{\mu},\delta) = 1 - \vec{c}e^{AT}e^{ \delta S_\mu T}r_{0}$. The log-sensitivity is then easily calculated as~\parencite{dorf}
\begin{equation}
    s(S_{\mu},T) = \left. \frac{1}{e(T)}\frac{\partial \tilde{e}(T;S_{\mu},\delta)}{\partial \delta} \right|_{\delta = 0} = \left\lvert \frac{-1}{e(T)}\vec{c}e^{AT}(S_{\mu}T)r_0 \right\rvert.
\end{equation}
For each controller, we calculate $s(S_{\mu},T)$ for the same $1000$ dephasing operators used in the approach of Section~\ref{ss: mc} for $\mu \in \{1,2,\hdots , 1000\}$. To arrive at a single value of the log-sensitivity for each controller, and to maintain consistency with the KDE approach, we take the arithmetic mean over all dephasing operators and define
\begin{equation}
    s_{a}(S,T) = \frac{1}{1000} \sum_{\mu = 1}^{1000}s(S_{\mu},T)
\end{equation}
where the subscript $a$ indicates the log-sensitivity is derived from an analytic calculation of the perturbed trajectory of~\eqref{eq: perturbed_r}. 

\section{Analysis} \label{analysis}

We focus our analysis on three topics: the level of concordance between the two log-sensitivity measures $s_{k}(S,T)$ and $s_{a}(S,T)$, the degree to which the controllers in the data set exhibit robustness properties that align with the trade-off induced by the $S+T=I$ identity of classical feedback control, and the role played by input-output state orthogonal pairs in the fidelity and robustness of the controllers. For the first two categories we execute a pair of hypothesis tests based on the Kendall $\tau$ as a non-parametric measure of correlation and the Pearson $r$ as a measure of linear correlation. The analysis of the role of orthogonal pairs is based on a visual interpretation of the fidelity and robustness plots for controllers that render input-output states as orthogonal pairs and those that do not.  

\subsection{Hypothesis Test}\label{hypothesis}

In Section~\ref{ss: robustness_comparison} below we test the concordance between $s_a(S,T)$ and $s_k(S,T)$. We expect a positive correlation between the two metrics and test the level of concordance through a one-tailed hypothesis test with the right tail. We establish
\begin{itemize}
\item $H_0$: no correlation between $s_a(S,T)$ and $s_k(S,T)$;
\item $H_1$: positive correlation between $s_a(S,T)$ and $s_{k}(S,T)$.
\end{itemize}
In Section~\ref{ss: trend_analysis} we compare the trend between each measure of log-sensitivity ($s_a(S,T)$ and $s_k(S,T)$) with the nominal error $e(T)$. The classical control trade-offs require a negative correlation between $e(T)$ and the log-sensitivity, so we establish a one-tailed hypothesis test on the left tail. For this test: 
\begin{itemize}
    \item $H_0$: no correlation between $e(T)$ and $\{s_a(S,T),s_k(S,T) \}$;
    \item $H_1$: negative correlation between $e(T)$ and $\{s_a(S,T),s_k(S,T)\}$.
\end{itemize}
In both cases we consider the $N=5$ transfers from $\ket{\IN = 1}$ to $\ket{\OUT} = \ket{2,3}$ and $N = 6$ transfers from $\ket{\IN} = \ket{1}$ to $\ket{\OUT} = \ket{2, 3, 4}$. For each of these five transfers we consider the controllers optimized for fidelity, dephasing, and overlap. With $100$ controllers within each transfer-optimization target combination we thus have $15$ tests for each hypothesis above, each with $100$ samples. 

To execute the computation of the Kendall $\tau$ and Pearson $r$ we leverage the MATLAB function \texttt{corr($\cdot,\cdot$)} with the option \texttt{'Kendall'} or \texttt{'Pearson'} as appropriate. In the following discussion of hypothesis tests, use of $\tau$ refers to the Kendall rank correlation coefficient and $r$ to the Pearson correlation coefficient. We compute the test statistic for the Kendall $\tau$ as $Z_{\tau} = \tau \left( \sqrt{\frac{2(2n+5)}{9n(n-1)}} \right)^{-1}$~\parencite{Kendall_tau_significance} where $n=100$ denotes the number of samples. We then determine the statistical significance of the test by evaluating 
\begin{equation}
    p_{\tau} = \begin{cases} 1 - \Phi(Z_{\tau}),& \text{ for }  s_k(S,T)  \text{ vs. } s_a(S,T), \\
    \Phi(Z_{\tau}),& \text{ for } \{s_a(S,T),s_k(S,T)\} \text{ vs. } e(T),
    \end{cases}
\end{equation}
where $\Phi(\cdot)$ is the normal cumulative distribution function. For the Pearson $r$-based test, we calculate the test statistic as $t_{r} = r \left( \sqrt{\frac{1-r^{2}}{n-2}}\right)^{-1}$. Here again, $n=100$. We then quantify the statistical significance of the test for a given value of $r$ as 
\begin{equation}
p_{r} = 
\begin{cases}
1- \mathcal{S}(t_{r}) , & s_k(S,T)  \text{ vs. } s_a(S,T),  \\
\mathcal{S}(t_r), & \{s_a(S,T),s_k(S,T)\} \text{ vs. } e(T),
\end{cases}
\end{equation}
where $\mathcal{S}$ represents the cumulative Student's $t$-distribution. 

Finally, we establish the level of significance at $\alpha = 0.02$ so that the hypothesis test itself is
\begin{itemize}
\item accept $H_0$ if $p_{\tau,r} \geq \alpha$,
\item reject $H_0$, if $p_{\tau,r} < \alpha$,
\end{itemize}
for both tests.

\subsection{Comparison of Robustness Assessments}\label{ss: robustness_comparison}

\begin{figure}[ht]
\includegraphics[
width = 1\textwidth]{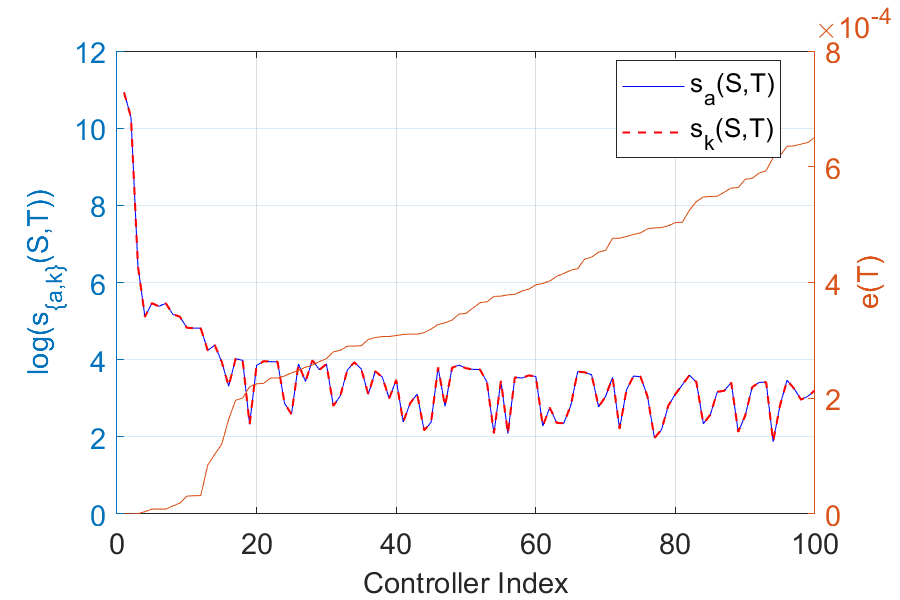}
\caption{Comparison $s_{a}(S,T)$ (solid blue line) vs.\ $s_{k}(S,T)$ (dashed red line) for fidelity-optimized controllers in an $N=6$ ring with transfer to spin $3$. Note the near perfect agreement of the two log-sensitivity measures. We can also observe here the conventional trend between both log-sensitivity measures and the fidelity error indicating a trade-off between performance (low $e(T)$) and robustness (small $s_{\{a,k\}}(S,T)$). } \label{fig: comp_6_3_fidelity}
\end{figure}

The hypothesis test reveals strong agreement between $s_{a}(S,T)$ and $s_{k}(S,T)$ for all $15$ tests with a Kendall $\tau$ or Pearson $r$ of near unity in all cases. Table~\ref{t: table_1} summarizes the results for the Kendall $\tau$-based hypothesis test. As depicted, all $15$ test cases result in rejection of $H_0$ in support of $H_1$ with $p$-values near zero. The Pearson $r$-based test provides similarly strong evidence of linear correlation. Though not indicative of equality, the hypothesis test confirms strong concordance. Visually, we can see this strong correlation in Figure~\ref{fig: comp_6_3_fidelity}, where both measures lie nearly on-top of the other.  While the analytic calculation is the preferred method of assessing robustness, evidence of this correlation supports the efficacy of the kernel density estimation measure when the equations of motion are so complex that available computing power makes this Monte Carlo approach more efficient than an analytic calculation.
  
\begin{table}[t]
    \centering
    \begin{tabular}{|l|l|l|l|l|}
    \hline
        \multicolumn{5}{|c|}{$s_{a}(S,T)$ vs. $s_{k}(S,T)$} \\ \hline
        Controller Type & Transfer & $\tau$ & $Z_{\tau}$ & $p$ \\ \hline
        Dephasing & N=5 out=2  & 1.000 & 14.742 & 0.000 \\ \hline
        Dephasing & N=5 out=3 & 1.000 & 14.736 & 0.000 \\ \hline
        Dephasing & N=6 out=2 & 1.000 & 14.742 & 0.000 \\ \hline
        Dephasing & N=6 out=3 & 1.000 & 14.742 & 0.000 \\ \hline
        Dephasing & N=6 out=4 & 1.000 & 14.742 & 0.000 \\ \hline
        Fidelity & N=5 out=2  & 1.000 & 14.742 & 0.000 \\ \hline
        Fidelity & N=5 out=3 & 1.000 & 14.742 & 0.000 \\ \hline
        Fidelity & N=6 out=2 & 1.000 & 14.736 & 0.000 \\ \hline
        Fidelity & N=6 out=3 & 1.000 & 14.736 & 0.000 \\ \hline
        Fidelity & N=6 out=4 & 1.000 & 14.742 & 0.000 \\ \hline
        Overlap & N=5 out=2  & 1.000 & 14.742 & 0.000 \\ \hline
        Overlap & N=5 out=3 & 1.000 & 14.742 & 0.000 \\ \hline
        Overlap & N=6 out=2 & 1.000 & 14.736 & 0.000 \\ \hline
        Overlap & N=6 out=3 & 1.000 & 14.742 & 0.000 \\ \hline
        Overlap & N=6 out=4 & 1.000 & 14.742 & 0.000 \\ \hline
    \end{tabular}
    \caption{Results of Kendall $\tau$-based hypothesis test for the concordance of $s_{a}(S,T)$ and $s_{k}(S,T)$. The hypothesis test provides strong confirmation that both the analytic and kernel density estimation are consistent in the evaluation of robustness to dephasing.}
    \label{t: table_1}
\end{table}

\subsection{Robustness Trend Analysis}\label{ss: trend_analysis}

\begin{figure}[t]
\subfloat[$5$-ring nearest-neighbor transfer with controllers optimized for fidelity\label{subfig:3a}]{\includegraphics[width = 1\textwidth]{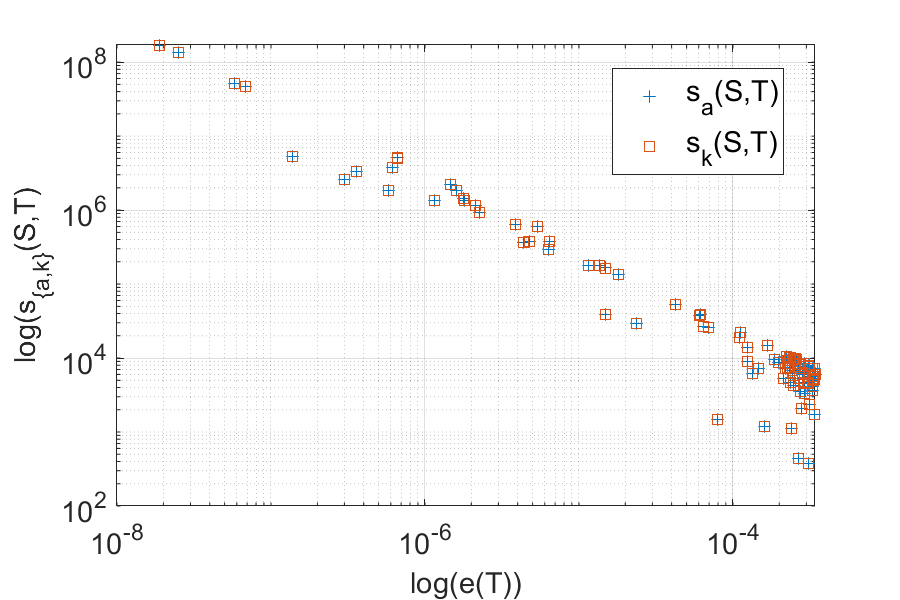}}
\hfill
\subfloat[$6$-ring nearest-neighbor transfer with controllers optimized for fidelity\label{subfig:3b}]{\includegraphics[width = 1\textwidth]{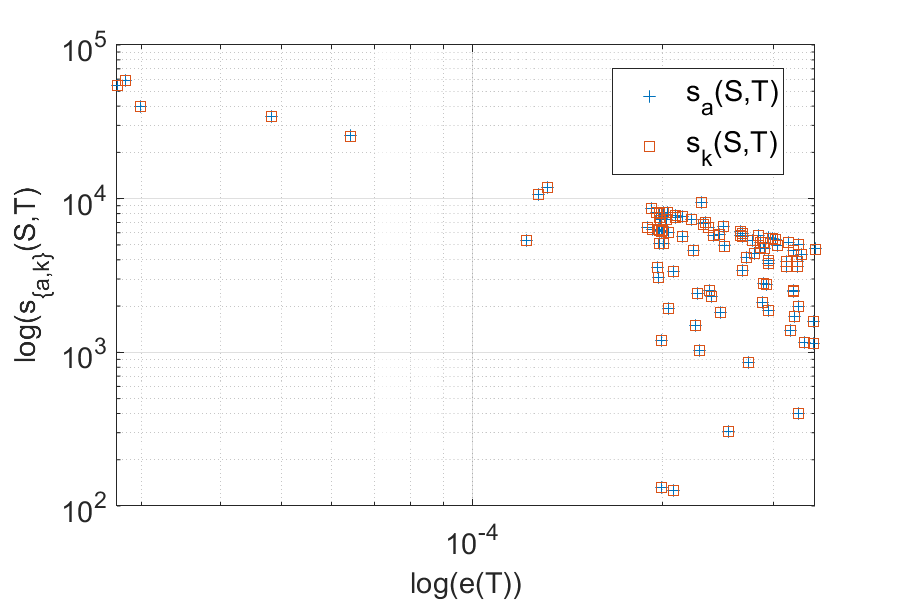}}
\caption{$s_{a}(S,T)$ and $s_{k}(S,T)$ versus $e(T)$ on a $\log-\log$ scale. Plot~\ref{subfig:3a} suggests a strong linear correlation, which is confirmed by a Pearson $r$ of $-0.972$, while~\ref{subfig:3b} shows a much weaker linear correlation with $r = -0.589$.} \label{fig: comparison}
\end{figure}

The results of the hypothesis test to evaluate concordance of the log-sensitivity with the fidelity error are summarized in Table~\ref{t: table_2}. The table presents the results of the Pearson $r$-based test, which evaluates the level of linear correlation between the metrics on a $\log-\log$ scale. For all $15$ test cases we see rejection of $H_0$ in favor of $H_1$ with $p$-values near zero. This suggests a strong negative correlation between the two metrics. Furthermore, the Pearson $r$ provides the slope of the best linear fit through the data point, valuable for assessing the impact of a given change in error on robustness. Figure~\ref{subfig:3a} shows a plot of $s_{a}(S,T)$ and $s_{k}(S,T)$ versus $e(T)$ on a log-log scale for a $5$-ring, nearest-neighbor transfer and fidelity-optimized controller. The near unity slope of the plot and $r$ value of $-0.972$ indicates a nearly uniform cost in robustness, measured by the log-sensitivity, for a given increase in performance, quantified by the fidelity error. Conversely, Figure~\ref{subfig:3b} depicts the correlation for controllers optimized for fidelity in a $6$-ring for nearest neighbor transfer. The linear correlation coefficient is less strong than that in Figure~\ref{subfig:3a} with $r = -0.5882$, indicating a less stringent adherence to a uniform cost in robustness for increased performance. The visual plot confirms this. For the highest fidelity controllers ($e(T) < 10^{-4}$ in the figure) we still observe a nearly linear trend. However, in the right-hand side of the plot ($e(T) > 10^{-4}$ in the figure), we observe a large number of controllers with varying log-sensitivity for the same error. 

\begin{table*}[t]
    \centering
    \begin{tabular}{|l|l|c|c|c|c|c|c|}
    \hline
        \multicolumn{2}{|c|}{} & \multicolumn{3}{c|}{$\log[s_{k}(S,T)]$ vs. $\log[e(T)]$} & \multicolumn{3}{c|}{$\log[s_{a}(S,T)]$ vs. $\log[e(T)]$} \\ \hline
        Controller Type & Transfer & $r$ & $t_r$ & $p_r$ & $r$ & $t_r$ & $p_r$ \\ \hline
        Dephasing & N=5 out=2  & -0.9576 & -32.8892 & 0.0000 & -0.9575 & -32.8548 & 0.0000 \\ \hline
        Dephasing & N=5 out=3 & -0.7262 & -10.4582 & 0.0000 & -0.7259 & -10.4493 & 0.0000 \\ \hline
        Dephasing & N=6 out=2 & -0.6512 & -8.4956 & 0.0000 & -0.6510 & -8.4908 & 0.0000 \\ \hline
        Dephasing & N=6 out=3 & -0.9193 & -23.1170 & 0.0000 & -0.9192 & -23.1045 & 0.0000 \\ \hline
        Dephasing & N=6 out=4 & -0.8630 & -16.9128 & 0.0000 & -0.8629 & -16.9029 & 0.0000 \\ \hline
        Fidelity & N=5 out=2  & -0.9723 & -41.1452 & 0.0000 & -0.9722 & -41.1000 & 0.0000 \\ \hline
        Fidelity & N=5 out=3 & -0.8684 & -17.3407 & 0.0000 & -0.8683 & -17.3271 & 0.0000 \\ \hline
        Fidelity & N=6 out=2 & -0.5882 & -7.1994 & 0.0000 & -0.5880 & -7.1960 & 0.0000 \\ \hline
        Fidelity & N=6 out=3 & -0.9160 & -22.5967 & 0.0000 & -0.9160 & -22.5995 & 0.0000 \\ \hline
        Fidelity & N=6 out=4 & -0.8498 & -15.9608 & 0.0000 & -0.8497 & -15.9504 & 0.0000 \\ \hline
        Overlap & N=5 out=2  & -0.6895 & -9.4229 & 0.0000 & -0.6893 & -9.4182 & 0.0000 \\ \hline
        Overlap & N=5 out=3 & -0.9410 & -27.5300 & 0.0000 & -0.9409 & -27.4905 & 0.0000 \\ \hline
        Overlap & N=6 out=2 & -0.4157 & -4.5253 & 0.0000 & -0.4156 & -4.5238 & 0.0000 \\ \hline
        Overlap & N=6 out=3 & -0.8907 & -19.3993 & 0.0000 & -0.8905 & -19.3802 & 0.0000 \\ \hline
        Overlap & N=6 out=4 & -0.8836 & -18.6857 & 0.0000 & -0.8835 & -18.6727 & 0.0000 \\ \hline
    \end{tabular}
    \caption{Table summarizing the Pearson $r$-based hypothesis test results for log-sensitivity versus $e(T)$. The test results in rejection of $H_0$ in favor of $H_1$ in all cases, indicating a trend in agreement with the conventional trade-off between performance and robustness.}
    \label{t: table_2}
\end{table*}

\subsection{Orthogonal Pair States, Robustness, and Fidelity} \label{ss: orthogonal states}

Given the suitability of orthogonal pairs to maximize the combined objective of fidelity under unitary transfer with asymptotic transfer fidelity, we expect that controllers which render the input and output states orthogonal pairs will dominate the ``overlap'' controller data set. Of more interest, however, are what fidelity and robustness properties input-output orthogonal pairs present across the breadth of the controllers. As Table~\ref{table_orthogonal} confirms, input-output states that form orthogonal pairs produce the best (highest-fidelity) controllers for the case of overlap-optimization. For both $5$ and $6$ rings, all overlap-optimized controllers for nearest-neighbor and next-nearest-neighbor (i.e.\ $\ket{OUT} = \ket{3}$) transfers create an orthogonal-pair input and output state. However, for the $N=6$ and $1 \rightarrow 4$ transfer, the population of overlap-optimized controllers is almost evenly split. For both fidelity-optimized and dephasing-optimized controllers, the controllers that render input-output orthogonal pairs dominate the nearest-neighbor transfers for both $N=5$ and $N=6$. However, for all other transfers, orthogonal pairs constitute a minority of the controllers. Given that the controllers under consideration were filtered for the best performance (fidelity), this suggests that controllers facilitating the highest levels of fidelity for nearest-neighbor transfer, \emph{regardless of optimization choice}, exhibit this orthogonal-pair property. This suggests optimizing to produce input-output eigenstructures that replicate orthogonal pairs as a means to generate high fidelity controllers under a range of conditions for nearest neighbor transfer.

\begin{table}[t]
    \centering
    \caption{Table depicting the percentage of input-output states rendered as orthogonal-pairs by each type of controller and each transfer.}
    \begin{tabular}{|l|l|c|c|}
    \hline
        Controller Type & Transfer & \% Orthogonal & \% Non-orthogonal \\
         & & Pairs & Pairs \\ \hline
        Dephasing & N=5 out=2  & 81 & 19 \\ \hline
        Dephasing & N=5 out=3 & 41 & 59 \\ \hline
        Dephasing & N=6 out=2 & 98 & 2 \\ \hline
        Dephasing & N=6 out=3 & 27 & 73 \\ \hline
        Dephasing & N=6 out=4 & 26 & 74 \\ \hline
        Fidelity & N=5 out=2  & 77 & 23 \\ \hline
        Fidelity & N=5 out=3 & 39 & 61 \\ \hline
        Fidelity & N=6 out=2 & 97 & 3 \\ \hline
        Fidelity & N=6 out=3 & 24 & 76 \\ \hline
        Fidelity & N=6 out=4 & 28 & 72 \\ \hline
        Overlap & N=5 out=2  & 100 & 0 \\ \hline
        Overlap & N=5 out=3 & 100 & 0 \\ \hline
        Overlap & N=6 out=2 & 100 & 0 \\ \hline
        Overlap & N=6 out=3 & 100 & 0 \\ \hline
        Overlap & N=6 out=4 & 47 & 53 \\ \hline
    \end{tabular}
    \label{table_orthogonal}
\end{table}

\begin{figure}[ht]
\centering
\subfloat[$5$-ring, nearest neighbor transfer, dephasing-optimized controllers\label{subfig:4a}]{\includegraphics[width = 1\textwidth]{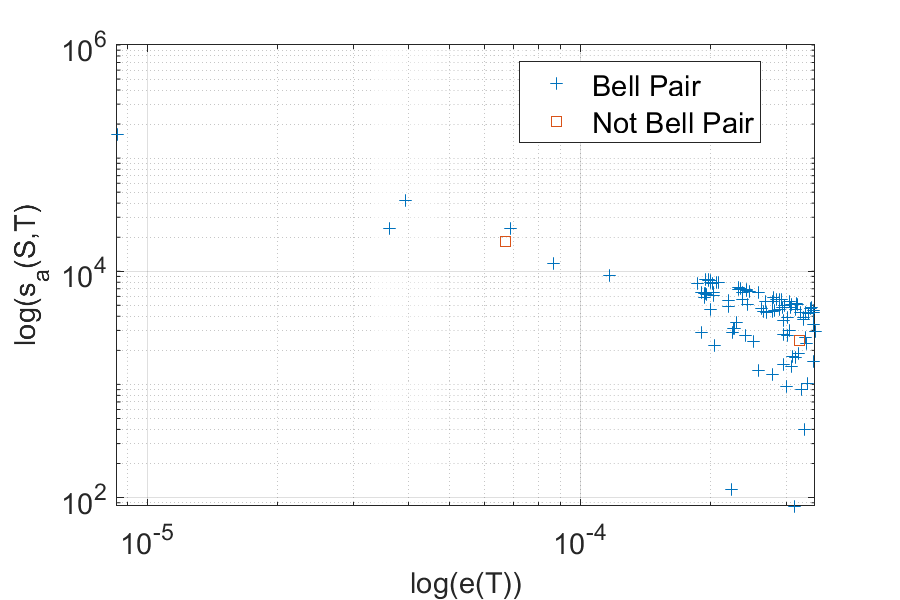}}
\hfill
\subfloat[$6$-ring, $\ket{1} \rightarrow \ket{4}$ transfer, fidelity-optimized controllers\label{subfig:4b}]{
\includegraphics[width = 1\textwidth]{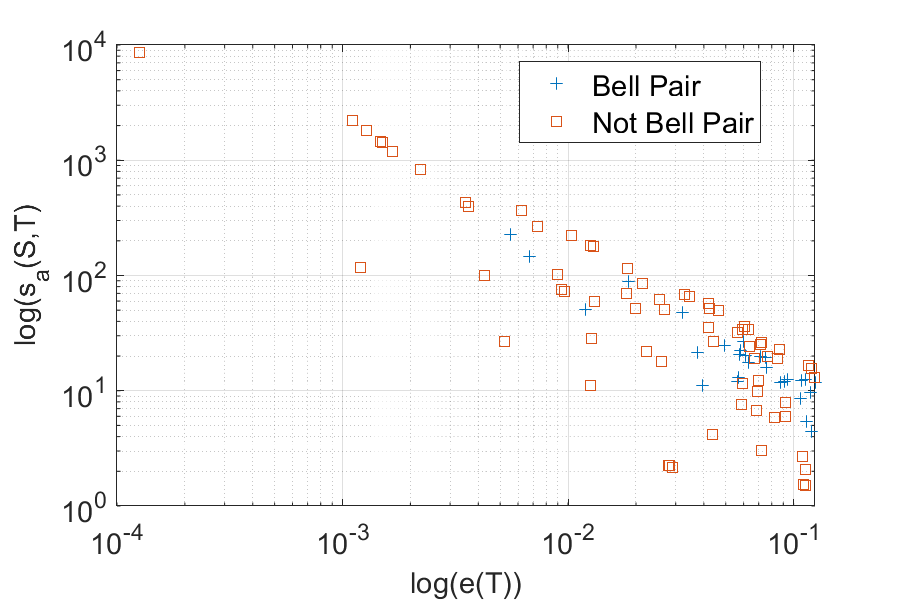}}
\caption{Plot of $s_{a}(S,T) $ vs.\ $e(T)$ on a $\log-\log$ scale showing controllers that yield orthogonal-pair input-output states in the eigenbasis of the Hamiltonian (blue crosses) and those that do not (red boxes). Orthogonal-pairs dominate, show the greatest fidelity, and also exhibit the smallest log-sensitivity in Figure~\ref{subfig:4a} while in Figure~\ref{subfig:4b} both the highest fidelity controllers and most robust controllers are not of the orthogonal-pair variety.}  
\label{fig: orthogonal_states}
\end{figure}

The robustness properties of the orthogonal-state controllers are less clear. For fidelity-optimized and dephasing-optimized controllers and nearest-neighbor transfer, the lower sensitivity controllers exhibit the orthogonal pair property as seen in Figure~\ref{subfig:4a}. Whereas for non-nearest-neighbor transfers, the more robust controllers appear to not render the input-output states orthogonal pairs. This behavior is evident in Figure~\ref{subfig:4b} where the most robust fidelity-optimized controllers in a $6$-ring for the $\ket{1} \rightarrow \ket{4}$ transfer are not of the orthogonal pair variety. Whether the robustness properties of the controllers in these cases are a characteristic of the orthogonal-like controllers or simply due to the dominance of orthogonal-type controllers in nearest-neighbor transfers still requires further investigation. 

\section{Conclusion} \label{conclusion}

In this paper we applied two distinct approaches to evaluate the log-sensitivity of the fidelity error to a perturbation of the system dynamics in the form of dephasing in the Hamiltonian basis. The KDE approach based on the error measurements of $1000$ dephasing operators introduced at varying strength and with the bandwidth input as described in Section~\ref{ss: mc} produced log-sensitivity values nearly identical to those of the analytic calculation based on the same dephasing operators. As distinct from previous work, we also considered controllers optimized for not only fidelity under unitary dynamics but fidelity under dephasing and fidelity in the asymptotic regime. We showed that in all cases the relationship between the fidelity error and the log-sensitivity adheres to the trade-off between performance and robustness expected from classical control theory. Of note, though we examined controllers optimized to maximize fidelity under dephasing, these controllers exhibited no better robustness to dephasing than the other optimization choices. This suggests that optimizing for fidelity under decoherence does not accrue robustness benefits, as measured by the log-sensitivity, over optimizing for fidelity under unitary transfer without exact knowledge of the dephasing process. 

Several items still require further investigation. Firstly, we only considered perturbations in the form of dephasing. It is necessary to generalize the analytic formula to allow for perturbations in the form of dephasing simultaneously with uncertainty in the Hamiltonian or control parameters as well as to extend the analysis to consider general decoherence processes that include dissipation with dephasing. This would facilitate a better understanding of robustness for general open quantum system, a persistent challenge within quantum control. Next, an investigation of the robustness properties of the controllers that produce input-output state orthogonal pairs is in order. In particular, though these types of controllers tend to provide the best fidelity for nearest-neighbor transfer, a comparison of their robustness properties with controllers that do not provide this orthogonal-pair property is missing, mainly due to the paucity of non-orthogonal type controllers for nearest-neighbor transfers for the controllers considered in this study. An investigation into this matter would provide a justification for either pursuing controllers of another sort, should they provide greater robustness for the same fidelity, or relying on orthogonal-type controllers to provide the best fidelity and robustness for nearest-neighbor transfer and optimizing with that target as the goal.

\section*{Conflict of interest}

The authors report no conflicts of interest.

\section*{Financial support}

Sean O'Neil acknowledges PhD funding through the US Army Advanced Civil Schooling program. 

\section*{Data availability}

The data is available at~\parencite{DataSet2_code} and~\parencite{DataSet2_results}.

\section{Connections references} 
Shermer, S. (2023). What is robust control in quantum technology? \textit{Research Directions: Quantum Technologies}, 1, e3, 1-3. \href{https://doi.org/10.1017/qut.2022.5}{https://doi.org/10.1017/qut.2022.5} 

\printbibliography

@article{Glaser2015,
  title={Training Schr{\"o}dinger’s cat: Quantum optimal control},
  author={Glaser, Steffen J and Boscain, Ugo and Calarco, Tommaso and Koch, Christiane P and K{\"o}ckenberger, Walter and Kosloff, Ronnie and Kuprov, Ilya and Luy, Burkhard and Schirmer, Sophie and Schulte-Herbr{\"u}ggen, Thomas and others},
  journal={The European Physical Journal D},
  volume={69},
  number={12},
  pages={279},
  year={2015},
  publisher={Springer Berlin Heidelberg}
}

@article{Edmond_IEEE_AC,
author={S. Schirmer and E. Jonckheere and F. Langbein},
title={Design of feedback control laws for information transfer in spintronics networks},
journal={IEEE Transactions on Automatic Control},
volume = {63},
number = {8},
pages = {2523-2536},
note={available at arXiv:1607.05294},
year={2018}
}

@inproceedings{time_optimal,
author={F. Langbein and S. Schirmer and E. Jonckheere},
title={Time optimal information transfer in spintronics networks},
booktitle={IEEE Conference on Decision and Control},
address={Osaka, Japan},
month={12},
year={2015},
pages={6454-6459}
}

@article{Schirmer2010,
  title={Stabilizing open quantum systems by Markovian reservoir engineering},
  author={Schirmer, SG and Wang, Xiaoting},
  journal={Physical Review A},
  volume={81},
  number={6},
  pages={062306},
  year={2010},
  publisher={APS}
}

@article{Ticozzi2010,
  title={Stabilizing quantum states by constructive design of open quantum dynamics},
  author={Ticozzi, Francesco and Schirmer, Sophie G and Wang, Xiaoting},
  journal={Automatic Control, IEEE Transactions on},
  volume={55},
  number={12},
  pages={2901--2905},
  year={2010},
  publisher={IEEE}
}

@article{Motzoi2016,
  title={Backaction-driven, robust, steady-state long-distance qubit entanglement over lossy channels},
  author={Motzoi, Felix and Halperin, Eli and Wang, Xiaoting and Whaley, K Birgitta and Schirmer, Sophie},
  journal={Physical Review A},
  volume={94},
  number={3},
  pages={032313},
  year={2016},
  publisher={American Physical Society}
}

@article{statistical_control,
title={Jonckheere-{T}erpstra test for nonclassical error versus log-sensitivity relationship of quantum spin network controllers},
author={E. Jonckheere and S. Schirmer and F. Langbein},
year={2018},
journal={International Journal of Robust and Nonlinear Control},
note={in press, available at  	arXiv:1612.02784 [math.OC]}
}

@inproceedings{domenico_CDC,
author={D. D'Alessandro and E. Jonckheere and R. Romano},
title={Control of open quantum systems in a Bosonic bath},
booktitle={54th IEEE Conference on Decision and Control (CDC)},
address={Osaka, Japan},
month={12},
year={2015},
pages={6460-6465},
note={Available at {\tt http:eudoxus2.usc.edu}}
}

@inproceedings{singular_vs_weak_coupling,
author={D. P. D'Alessandro and E. Jonckheere and R. Romano},
title={On the control of open quantum systems in the weak coupling limit},
booktitle={21st International Symposium on the Mathematical Theory of Networks and Systems ({MTNS})},
year={2014},
month={7},
address={Groningen, the Netherlands},
pages={1677-1684}
}

@article{neat_formula,
author={F. F. Floether and P. de Fouquieres and S. Schirmer},
title={Robust quantum gates for open systems via optimal control: {M}arkovian versus non-{M}arkovian dynamics},
journal={New Journal of Physics},
volume={14},
year={2012},
pages={1-26},
note={073023}
}

@article{Bertlmann_2008,
doi = {10.1088/1751-8113/41/23/235303},
url = {https://dx.doi.org/10.1088/1751-8113/41/23/235303},
year = {2008},
month = {may},
publisher = {},
volume = {41},
number = {23},
pages = {235303},
author = {Reinhold A Bertlmann and Philipp Krammer},
title = {Bloch vectors for qudits},
journal = {Journal of Physics A: Mathematical and Theoretical}
}

@incollection{Kendall_tau_significance,
author={H. Abdi},
title={The {K}endall rank correlation coefficient},
booktitle={Encyclopedia of measurements and Statistics},
editor={N. Salkind},
publisher={Sage},
address={Thousand Oaks, CA},
year={2007}
}

@article{PhysRevA.86.012121,
  title = {Limits on the decay rate of quantum coherence and correlation},
  author = {Oi, Daniel K. L. and Schirmer, Sophie G.},
  journal = {Phys. Rev. A},
  volume = {86},
  issue = {1},
  pages = {012121},
  numpages = {7},
  year = {2012},
  month = {7},
  publisher = {American Physical Society},
  doi = {10.1103/PhysRevA.86.012121},
  url = {https://link.aps.org/doi/10.1103/PhysRevA.86.012121}
}

@article{Awschalom2013,
   author = {David D Awschalom and Lee C Bassett and Andrew S Dzurak and Evelyn L Hu and Jason R Petta},
   doi = {10.1126/science.1231364},
   issue = {6124},
   journal = {Science},
   pages = {1174-1179},
   title = {Quantum Spintronics: Engineering and Manipulating Atom-Like Spins in Semiconductors},
   volume = {339},
   url = {http://www.sciencemag.org/content/339/6124/1174.abstract},
   year = {2013},
}

@article{Schirmer2022,
   author = {Sophie G. Schirmer and Frank C. Langbein and Carrie Ann Weidner and Edmond Jonckheere},
   doi = {10.1109/TAC.2022.3181249},
   issn = {0018-9286},
   issue = {11},
   journal = {IEEE Transactions on Automatic Control},
   month = {11},
   pages = {6012-6024},
   title = {Robust Control Performance for Open Quantum Systems},
   volume = {67},
   year = {2022},
}

@article{Altafini2012,
   author = {Claudio Altafini and Francesco Ticozzi},
   doi = {10.1109/TAC.2012.2195830},
   month = {10},
   title = {Modeling and Control of Quantum Systems: An Introduction},
   url = {http://arxiv.org/abs/1210.7127 http://dx.doi.org/10.1109/TAC.2012.2195830},
   year = {2012},
}

@misc{DataSet1,
   author = {Frank Langbein and Sean O'Neil and Sophie Shermer},
   city = {Cambridge, UK},
    doi = {https://doi.org/10.33774/coe-2022-35xgg},
   publisher = {Cambridge University Press},
   title = {Energy landscape controllers for quantum state transfer in spin-1/2 networks with ring topology},
   url = {https://www.cambridge.org/engage/coe/article-details/63c9eda607308a0e4c9f89a9?show=item},
   year = {2022},
}

@inproceedings{CDC2018,
   author = {S. Schirmer and E. Jonckheere and S. O'Neil and F. C. Langbein},
   doi = {10.1109/CDC.2018.8619179},
   isbn = {978-1-5386-1395-5},
   journal = {2018 IEEE Conference on Decision and Control (CDC)},
   month = {12},
   pages = {6608-6613},
   publisher = {IEEE},
   title = {Robustness of Energy Landscape Control for Spin Networks Under Decoherence},
   url = {https://ieeexplore.ieee.org/document/8619179/},
   year = {2018},
}

@misc{oneil_2022,
  doi = {10.48550/ARXIV.2210.15783}, 
  url = {https://arxiv.org/abs/2210.15783},
  author = {O'Neil, S. and Schirmer, S. G. and Langbein, F. C. and Weidner, C. A. and Jonckheere, E.},
  title = {Time Domain Sensitivity of the Tracking Error},
  publisher = {arXiv},
  year = {2022},
  copyright = {arXiv.org perpetual, non-exclusive license}
}

@article{Joel_2013,
	doi = {10.1119/1.4798343},
	url = {https://doi.org/10.1119%2F1.4798343},
	year = 2013,
	month = {6},
	publisher = {American Association of Physics Teachers ({AAPT})},
	volume = {81},
	number = {6},
	pages = {450--457},
	author = {Kira Joel and Davida Kollmar and Lea F. Santos},
	title = {An introduction to the spectrum, symmetries, and dynamics of spin-1/2 Heisenberg chains},
	journal = {American Journal of Physics}
}

@article{deph_rates,
  title = {Constraints on relaxation rates for $N$-level quantum systems},
  author = {Schirmer, S. G. and Solomon, A. I.},
  journal = {Phys. Rev. A},
  volume = {70},
  issue = {2},
  pages = {022107},
  numpages = {13},
  year = {2004},
  month = {8},
  publisher = {American Physical Society},
  doi = {10.1103/PhysRevA.70.022107},
  url = {https://link.aps.org/doi/10.1103/PhysRevA.70.022107}
}

@misc{DataSet2_code,
   author = {Frank Langbein and Sean O'Neil and Sophie Shermer},
   city = {Cambridge, UK},
   title = {Data - Energy Landscape Controllers},
   url= {https://qyber.black/spinnet/data-elc-xx-rings},
   year = {2022},
}

@misc{DataSet2_results,
   author = {Frank Langbein and Sean O'Neil and Sophie Shermer},
   city = {Cambridge, UK},
   title = {Results - Robustness of Energy Landscape Controllers},
   url= {https://qyber.black/spinnet/results-elc-xx-rings-robustness},
   year = {2022},
}

@article{bio-imaging,
author = {Chi-Cheng Fu  and Hsu-Yang Lee  and Kowa Chen  and Tsong-Shin Lim  and Hsiao-Yun Wu  and Po-Keng Lin  and Pei-Kuen Wei  and Pei-Hsi Tsao  and Huan-Cheng Chang  and Wunshain Fann },
title = {Characterization and application of single fluorescent nanodiamonds as cellular biomarkers},
journal = {Proceedings of the National Academy of Sciences},
volume = {104},
number = {3},
pages = {727-732},
year = {2007},
doi = {10.1073/pnas.0605409104},
URL = {https://www.pnas.org/doi/abs/10.1073/pnas.0605409104},
eprint = {https://www.pnas.org/doi/pdf/10.1073/pnas.0605409104}
}

@article{
quantum_dots,
author = {F. Pelayo García de Arquer  and Dmitri V. Talapin  and Victor I. Klimov  and Yasuhiko Arakawa  and Manfred Bayer  and Edward H. Sargent },
title = {Semiconductor quantum dots: Technological progress and future challenges},
journal = {Science},
volume = {373},
number = {6555},
pages = {eaaz8541},
year = {2021},
doi = {10.1126/science.aaz8541},
URL = {https://www.science.org/doi/abs/10.1126/science.aaz8541},
eprint = {https://www.science.org/doi/pdf/10.1126/science.aaz8541}
}

@article{Koch_2022,
doi = {10.1140/epjqt/s40507- 022-00138-x},
url = {https://doi.org/10.1140\%2Fepjqt\%2Fs40507-022-00138-x},
year = 2022,
month = {7},
publisher = {Springer Science and Business Media {LLC}
},  
volume = {9},
  number = {1},
  author = {Christiane P. Koch and Ugo Boscain and Tommaso Calarco and Gunther Dirr and Stefan Filipp and Steffen J. Glaser and Ronnie Kosloff and Simone Montangero and Thomas Schulte-Herbrüggen and Dominique Sugny and Frank K. Wilhelm},
  title = {Quantum optimal control in quantum technologies. Strategic report on current status, visions and goals for research in Europe},
journal = {{EPJ} Quantum Technology}
}

@article{scott_1979,
    author = {David W. Scott},
    title = "{On optimal and data-based histograms}",
    journal = {Biometrika},
    volume = {66},
    number = {3},
    pages = {605-610},
    year = {1979},
    month = {12},
    issn = {0006-3444},
    doi = {10.1093/biomet/66.3.605},
    url = {https://doi.org/10.1093/biomet/66.3.605},
    eprint = {https://academic.oup.com/biomet/article-pdf/66/3/605/632347/66-3-605.pdf},
}

@article{splines,
    author = {A. Perperoglou and W. Sauerbrei and M. Abrahamowicz and M. Schmid},
    title = {A review of spline function procedures in R.},
    journal = {BMC Med Res Methodol.},
    year = {2019},
    month = {3},
    volume = {19}, 
    number = {1}, 
    doi = {10.1186/s12874-019-0666-3},
    }

@article{data_set_1_analysis, 
title={Robustness of energy landscape controllers for spin rings under coherent excitation transport}, 
volume={1}, DOI={10.1017/qut.2023.5}, 
journal={Research Directions: Quantum Technologies}, author={O’Neil, Sean P. and Langbein, Frank C. and Jonckheere, Edmond and Shermer, Sophie}, 
year={2023}, 
pages={e12}
}

@book{dorf,
author = {Dorf, Richard C. and Bishop, Robert H.},
title = {Modern Control Systems},
year = {2000},
isbn = {0130306606},
publisher = {Prentice-Hall, Inc.},
address = {USA},
edition = {9th}
}

@inproceedings{Burhenne_2011,
author = {Burhenne, Sebastian and Jacob, Dirk and Henze, Gregor},
year = {2011},
month = {01},
pages = {1816-1823},
title = {Sampling based on Sobol' sequences for Monte Carlo techniques applied to building simulations},
journal = {Proceedings of Building Simulation 2011: 12th Conference of International Building Performance Simulation Association}
}

@article{Silverman_1986, 
title={Density Estimation for Statistics and Data Analysis}, 
DOI={10.1201/9781315140919}, 
journal={Routledge eBooks}, 
author={Silverman, Bernard W.}, 
year={1986}, 
month={Jan} 
}

@article{Gorini_1977,
title = {Properties of quantum Markovian master equations},
journal = {Reports on Mathematical Physics},
volume = {13},
number = {2},
pages = {149-173},
year = {1978},
issn = {0034-4877},
doi = {https://doi.org/10.1016/0034-4877(78)90050-2},
url = {https://www.sciencedirect.com/science/article/pii/0034487778900502},
author = {Vittorio Gorini and Alberto Frigerio and Maurizio Verri and Andrzej Kossakowski and E.C.G. Sudarshan}
}

\end{document}